\documentclass{aastex62}

\usepackage{amsmath}
\usepackage{amssymb}	

\usepackage{graphicx}
\usepackage[outdir=./]{epstopdf}

\graphicspath{{./}{figures/}}

\shorttitle{GRB's X-ray Flares}
\shortauthors{Shahamat et al.}

\begin{document}

\title{Viscous Evolution of Magnetized Clumps: a Source for X-ray Flares in Gamma-ray Bursts}

\email{n.shahamat.d@gmail.com, abbassi@um.ac.ir}

\author{Narjes Shahamat}
\affil{Department of Physics, School of Science, Ferdowsi University of Mashhad, Mashhad,PO Box 91775-1436 Iran}

\author{Shahram Abbassi}
\affil{Department of Physics, School of Science, Ferdowsi University of Mashhad, Mashhad,PO Box 91775-1436 Iran}
\affiliation{School of Astronomy , Institute for Studies in Theoretical Physics and Mathematics, PO Box 19395-5531, Tehran, Iran}




\begin{abstract}

X-ray flares can be accounted for a hint to the late time activity of Gamma-ray bursts (GRBs) central engines. Such a long term activity has been described through some models, one of which is the viscous evolution of the outer disc's fragments that proposed by \citet{2006ApJ...636..L29}, and developed quantitatively by \citet{2017MNRAS...464..4399}. Here, we reconstruct \citet{2017MNRAS...464..4399} framework through taking both small and large scale effects of magnetic field into account. To consider the magnetic barrier as a possible mechanism that might govern the accretion process of each magnetized clump, we make a simple pattern in boundary condition through which this mechanism might happen. Regarding various model parameters, we probe for their influence and proceed some key analogies between our model predictions and previous phenomenological estimates, for two different choices of boundary conditions (with and without a magnetic barrier). Our model is remarkably capable of matching flare bolometric and X-ray light-curves, as well as reproducing their statistical properties, such as the ratios between rise and decay time, width parameter and peak time, and the power-law correlation between peak luminosity and peak time. Combining our results with the conclusions of previous studies, we are led to interpret magnetic barrier as a less probable mechanism that might control the evolution of these clumps, especially the later created (or viscously evolved) ones.   

\end{abstract}



\section{Introduction} \label{sec1}

X-ray flares are detected in about one third of Swift gamma-ray bursts (GRBs) (\citet{2005Sci...309..183}; \citet{2006ApJ...641..1010}; \citet{2006ApJ...642..389}; \citet{2006ApJ...642..354}), and marked as one of the most common phenomena that reveals the late time erratic behavior of GRBs' central engines. Flares are detected in both long and short GRBs (\citet{2006A&A...454..113}; \citet{2006ApJ...641..1010}; \citet{2006A&A...450..59}; \citet{2011aMNRAS...410..1064}), and appear mostly in $10^{2}-10^{5}s$ time window (\citet{2010MNRAS...406..2113}; \citet{2016ApJS...224..20}), hence overlapping with the afterglow time-scale. 

Several models have proposed over the years to describe this late time flaring activity. The leading external shock scenario (e.g. \citet{1997ApJ...476..232}) failed to justify the temporal properties of X-ray flares. In particular, \citet{2007MNRAS...375..46} argued that inhomogeneities in the external shock prevent this model to reproduce the observational properties of the flares. On the other hand, conducting more detailed analogies between the temporal and spectral properties of the flares and those of prompt emissions (e.g. \citet{2010bMNRAS...406..2149}) favors the idea that flares and prompt pulses are common in their origin, that is to say they both trace the central engine activity (for more discussions see, e.g. \citet{2005ApJ...631..429}; \citet{2016MNRAS...457..L108}). Therefore, studying the origin of the flares might open a new window on GRBs' central engines.

A variety of efforts have been conducted to attribute the flares to central engine evolutions. \citet{2005ApJ...630..L113} suggested that fragmentation of a collapsing star and its subsequent accretion can lead to X-ray flare production. Moreover, ring-like fragmentation of the outer regions in a hyperaccreting disc might cause such a flaring activity, proposed by \citet{2006ApJ...636..L29}. From the magnetic point of view, some scenarios have been proposed that support the long term activity of the central engine, e.g. late time episodic accretion caused by a magnetic flux accumulation in the inner disc regimes, i.e. magnetic barrier, (\citet{2006MNRAS...370L..61}), and magnetic reconnection in a differentially rotating millisecond pulsar (\citet{2006Science...311..1127}).      

Motivated by the observed correlation between flare's duration and its arrival time (\citet{2006ApJ...647..1213}; \citet{2006Nature...440..164}), and the similarity in the distribution of waiting times of X-ray flares and prompt gamma-ray emissions, which might lead to the consideration of a similar physical source for both events (\citet{2015ApJ...801..57}), \citet{2017MNRAS...464..4399} developed the idea of the disc fragmentation in a more quantitative manner. They suggest that various flares with different arrival times (including both early and late time ones) might be attributed to the viscous spreading of different clumps created during the early or late time phase of central engine evolution. More specifically, they introduced offset time to be the time delay between GRB trigger and the time at which X-ray flares set in, and argued that different offset times are subject to various clumps' viscous evolution starting point. So that the early flares are considered to be caused by the prompt accretion of clumps, generated during the early activity phase, while the delayed flares are due to the accreting clumps made further in time, or those of early generated fragments that need to migrate toward the inner regions for viscous stress to get dominated. They, then, provided a semi-analytical solution to this viscously evolved clump, and reproduced X-ray flare light curves interestingly well matched to the data, after implementing some spectral corrections. 

On the other hand, \citet{2017ApJ...845..64} studied the effects of magnetic field on both vertical structure and neutrino luminosity of a self-gravitating neutrino dominated accretion disc, as a plausible candidate for early phase activity of GRB's central engine. The strongly magnetized nature of the disc, with a magnetic field of about $\sim10^{15-16}G$, made both small and large scale effects of magnetic field (the former was regarded through magnetic viscosity, and the latter was studied via magnetic braking process) worth studying. They mainly found that such a consideration may shrink the gravitationally unstable regions toward the outer initial disc, and highlight magnetic barrier as a plausible mechanism for the lower accretion rates. All these findings together with the fact that GRB's central engines are supposed to be highly magnetized, lead us to develop \citet{2017MNRAS...464..4399} scenario through taking the effects of magnetic field (small and large scale) into account and probe how the combination of magnetic barrier and fragmentation might affect this intriguing model to describe X-ray flare's light curves. 

There are different approaches in the literature which addressed magnetic barrier mechanism. For instance, \citet{2010MNRAS...406..1208}, \citet{2011MNRAS...416..893} and \citet{2012MNRAS...420..416} studied a magnetically truncated disc around a star (such as neutron star). As they pointed out, the strong magnetic field near star truncates the accretion disc. Due to angular momentum transfer between magnetic field and disc, one plausible state is that the disc is truncated outside the corotation radius $R_{c}=(GM_{*}/\Omega_{*}^{2})^{(1/3)}$  (at which frequency of the star's rotation equals Keplerian frequency), with $M_{*}$ and $\Omega_{*}$ are the mass and rotational frequency of the star. In this case a centrifugal barrier prevents the disc from accreting. Furthermore, the magnetic field lines form a "propeller regime". Such a topology of the field lines causes the gas inside magnetospheric radius, $R_{m}$ (the radius at which the accumulated magnetic field disrupts the accretion flow), to flow in freely along field lines toward the star's magnetic poles (\citet{2003PASJ...55..L69}). On the other hand, \citet{2003PASJ...55..L69} discussed physical conditions govern the Magnetically Arrested Discs (MAD) around black holes. They argue that MADs are considered to be those of accreting structures which are disrupted at $R_{m}$, by the accumulated poloidal magnetic field in the vicinity of black holes. For $R>R_{m}$ the flow is axisymmetric while for $R<R_{m}$ the flow breaks up into blobs or streams. Such streams or blobs can flow in slowly (with a velocity much less than the free fall velocity) towards the black hole via magnetic interchanges and reconnection. Regarding physical conditions imposed by magnetic barrier around black holes, we adopt \cite{2003PASJ...55..L69} strategy which leads us to consider two limitations in boundary conditions. The first one is zero mass flux rate at the inner radius, and the second one is taking the magnrtospheric radius ($R_{m}$) as the inner boundary of the disc, namely $R_{in}$.   

In the present paper, section (\ref{sec2}) manifests our model framework with a clarification of all assumptions and approximations we made in order to smooth our way in extracting a semi-analytical solution, following \citet{2017MNRAS...464..4399} approach. The correlation between our model parameters and those related to the light curve shape, beside some key analogies of our model predictions with observations will be conducted in section (\ref{sec3}). We then summarize and discuss our main conclusions in section (\ref{sec4}).

\section{Ring-like fragments and their viscous evolution} \label{sec2}

Hyper-accreting discs are gravitationally unstable in their outer regions that might result in disc fragmentation and be accounted as a source of disc's late time activity (i.e., X-ray flares), as suggested by \citet{2006ApJ...636..L29}, and discussed by \citet{2014ApJ...791..69} and \citet{2017ApJ...845..64}. Disc gets gravitationally unstable once the Toomre parameter, $Q$, reads (\citet{1964ApJ...139..1217})

\begin{equation}
Q=\frac{c_{s}\Omega}{\pi G \Sigma}<1
\end{equation}  
where $c_{s}$ is the sound speed, $\Omega_{k}$ the local Keplerian angular velocity, $G$ is the gravitational constant and $\Sigma$ is the disc surface density. In magnetized case, this criterion takes the form (\citet{Shu1992})

 \begin{equation}
 Q_{mag}=\frac{\sqrt{c_{s}^{2}+v_{A}^{2}}\Omega}{\pi\Sigma G}<1
 \label{2}
 \end{equation}
 where $v_{A}=\frac{B}{\sqrt{4\pi \rho}}$ is the Alfven velocity. However, in these unstable regimes, fragmentation into bound objects will occur if (\citet{2001ApJ...553..174}; \citet{2006ApJ...636..L29})
 
  \begin{equation}
  t_{cool}<t_{\rm cirt}\approx 3\Omega^{-1},
  \end{equation}
  where cooling timescale is denoted by $t_{cool}\approx(H/R)^{2}t_{\nu}$ (\citet{1991MNRAS...248..754}), with $t_{\nu}=\frac{2}{3}\frac{R^{2}}{\nu}$ regarded as viscous time scale. Fragments merge and/or accrete until their tidal influence gets strong enough to open a gap in the disc. This happens when the mass of the clump has increased to
  
  \begin{equation}
  M_{frag}\simeq(\frac{H}{R})^{2}\alpha^{1/2}M_{BH}.
  \label{4}
  \end{equation}
   where $M_{BH}$ is the mass of the central black hole (\citet{1996ApJ...460..832}; \citet{2017MNRAS...464..4399}), $H$ is the half thickness of the disc, and $\alpha$ is the viscous parameter. 
   
   In what follows, regarding the model proposed by \citet{2006ApJ...636..L29}, we suppose such a ring-like clump has been created in the outer regions as a sharp accumulated mass, more specifically, a delta function, at radius $R_{0}$. Computing the viscous evolution of the clump, we also consider the standard equation for axisymmetric accretion discs (\citet{Kato2008})
   
   \begin{equation}
   \frac{\partial}{\partial t}\Sigma (R,t)=\frac{1}{R}\frac{\partial}{\partial R}[R^{1/2}\frac{\partial}{\partial R}(3\nu\Sigma R^{1/2})]
   \label{5}
   \end{equation} 
   
   Although this equation seems to ignore magnetic field effects, the small scale effect of magnetic field can be regarded through magnetic viscosity term, i.e. $\nu$. Hence, being interested in considering the large scale effect of magnetic field, we add a term corresponding to the torque exerted by the Lorentz force. To do so, we made use of \citet{2000Phys. Rev...325..83} and \citet{2017ApJ...845..64} approach to modify equation (\ref{5}) to
   
   \begin{equation}
   \frac{\partial}{\partial t}\Sigma (R,t)=-\frac{1}{R}\frac{\partial}{\partial R}[\frac{\frac{\partial}{\partial R}(R^{3}\nu\Sigma \frac{d\Omega}{dr})+\frac{R^{2}B_{\phi}B_{z}}{2\pi}}{\frac{d}{dR}(R^{2}\Omega)}]
   \label{6}
   \end{equation} 
   
   Considering the common assumption $B_{z}\approx B_{R}$ (e.g. \citet{2017ApJ...845..64}; \citet{2006MNRAS...370L..61}), in order to get rid of the magnetic field components in the second term of the numerator, we can make use of the magnetic viscosity equation
   \begin{equation}
   \frac{B_{R}B_{\phi}}{4\pi}=-\frac{3}{2}\alpha P,
   \label{7}
   \end{equation} 
   where $P$ is the total pressure of the accretion flow. On the other hand, we know that
   
  \begin{equation}
  P=-\frac{1}{\alpha}\rho\nu R\frac{d\Omega}{dR},
  \label{8}
  \end{equation}
  and consider $H\rho \approx \Sigma$, as a vertically averaged approximation (\citet{Kato2008}). Finally, one can rewrite equation (\ref{6}) as 
  
  \begin{equation}
  \frac{\partial}{\partial t}\Sigma (R,t)=-\frac{1}{R}\frac{\partial}{\partial R}[\frac{\frac{\partial}{\partial R}(R^{3}\nu\Sigma \frac{d\Omega}{dr})+3\frac{d\Omega}{dR}\frac{\Sigma}{h}\nu R^{2}}{\frac{d}{dR}(R^{2}\Omega)}],
  \label{9}
  \end{equation}
  where $h=\frac{H}{R}$. 
  
  In general, the viscosity $\nu$ depends on the surface density and equation (\ref{9}) is non-linear. If, however, $\nu$ is only a function of radius, then the equation is linear and much more amenable to analytic methods. Therefore, following \citet{2017MNRAS...464..4399} and \citet{2011MNRAS...410..1007}, we adopt the viscosity to have a radial power law, $\nu\propto R^{n}$, to achieve an exact solution for $\Sigma (R,t)$ using a Green's function $G$,
\begin{equation}
\Sigma (R,t)=\int_{R_{in}}^{\infty}G(R,R^{\prime},t)\Sigma(R^{\prime},t=0) dR^{\prime},\label{10}
\end{equation}
in which $\Sigma(R, t = 0)$ is a given arbitrary profile at $t=0$. Having $\Sigma (R,t)$, the accretion power $L_{acc}$, due to the viscous spreading of the clump, approximately reads

 \begin{equation}
 L_{acc}\simeq\int_{R_{in}}^{\infty}\frac{9}{4}\Sigma(R,t)\nu(R)\Omega^{2}(R)2\pi RdR.
 \label{11}
 \end{equation}

 
To compute Green's function, we need to determine our desired boundary condition. In this regard, we encounter two strategies. The first is zero central torque boundary condition, which is of astrophysical interest especially in case of accretion on to a black hole or a slowly rotating star, at radii larger than the radius of innermost circular orbit or the stellar surface, respectively. Another one is zero mass flux boundary condition. Having a strong central source of angular momentum, the accretion gas will be prevented to flow in, and instead accumulate near the center (\citet{2011MNRAS...410..1007}). Such solutions can describe accretion discs around a compact binary (\citet{1991MNRAS...248..754}), and compact objects with strong central magnetic fields (\citet{1974MNRAS...168..603}). The latter seems similar to the physical situation imposed by the accumulated poloidal magnetic field near the innermost region of magnetically arrested discs, which might result in magnetic barrier mechanism (\citet{2003PASJ...55..L69}; \citet{2017ApJ...845..64}). Therefore, we are to apply both boundary conditions, in order to probe the effect of magnetic barrier on the X-ray flare light curve.  

Obtaining the appropriate Green's function, in case of zero torque boundary condition, the surface density integral takes the following form

	{\small \begin{eqnarray}
		\frac{\Sigma}{\Sigma_{0}}=\int_{0}^{\infty}\frac{R}{R_{0}}^{-n+(1-b_{1})/2}(\frac{R_{0}}{R_{in}})^{2-n}(1-n/2)e^{-2\kappa^{2}(1-n/2)^2t/t_{\nu_{in}}}[Y_{l}(\kappa)J_{l}(\kappa x_{0})\nonumber\\\nonumber\\
		-Y_{l}(\kappa x_{0})J_{l}(\kappa)]\frac{[Y_{l}(\kappa)J_{l}(\kappa x)-Y_{l}(\kappa x)J_{l}(\kappa)]}{[Y^{2}_{l}(\kappa)+J_{l}^{2}(\kappa)]}\kappa d\kappa
		\label{12}
		\end{eqnarray}}
and in case of zero mass flux boundary condition one may achieve

	{\small \begin{eqnarray}
		\frac{\Sigma}{\Sigma_{0}}=\int_{0}^{\infty}\frac{R}{R_{0}}^{-n+(1-b_{1})/2}(\frac{R_{0}}{R_{in}})^{2-n}(1-n/2)e^{-2\kappa^{2}(1-n/2)^2t/t_{\nu_{in}}}[Y_{l-1}(\kappa)J_{l}(\kappa x_{0})\nonumber\\\nonumber\\
		-Y_{l}(\kappa x_{0})J_{l-1}(\kappa)]\frac{[Y_{l-1}(\kappa)J_{l}(\kappa x)-Y_{l}(\kappa x)J_{l-1}(\kappa)]}{[Y^{2}_{l-1}(\kappa)+J_{l-1}^{2}(\kappa)]}\kappa d\kappa
		\label{13}
		\end{eqnarray}}
To know about new parameters and variables, and get an intuition about the adopted procedure, we refer readers to appendix (\ref{appA}).

What matters right now, is how to fix the parameters $\Sigma_{0}$ and $R_{0}$.

\subsection{Determination of $\Sigma_{0}$ and $R_{0}$}
\label{sec2.1}

As previously stated, one can assume the ring-like clump as a sharp concentration at radius $R_{0}$. This radius can be regarded the same as the gravitational instability radius. Therefore, we approximated $\Sigma_{0}$ with the expression

\begin{equation}
\Sigma_{0}\simeq\frac{M_{frag}}{2\pi R_{0} l_{cl}}\approx\frac{M_{frag}}{2\pi R_{ins} l_{cl}}
\label{14}
\end{equation}
where $l_{cl}$ denotes the clump size, that can be estimated as the local Jeans length ($\lambda_{j}$). First, to evaluate $R_{0}$, we need to find an expression for $R_{ins}$. This can be fulfilled by the use of Toomre criterion (equation (\ref{2})). Through some mathematical considerations as performed by \citet{2017ApJ...845..64}, we can approximate magnetic pressure ($\frac{B^{2}}{8\pi}$) with the total pressure ($P$). Mathematicaly speaking, one knows that 
 \begin{equation*}
 B^{2}/8\pi<P+B^{2}/8\pi, 
  \end{equation*}
 we then have
  \begin{equation*}
  p_{mag}\approx\beta(P+B^{2}/8\pi)\ with\ \beta<1. 
  \end{equation*}
This provides us with the following expression 
  \begin{equation}
B^{2}=8\pi\frac{\beta}{1-\beta}P\approx 0.3P
\end{equation}
in which, to get the last equality, we choose $\beta$ to be $0.01$, which is the same value as adopted by \citet{2017ApJ...845..64}. With the help of equation (\ref{8}) we achieve
\begin{equation}
B^{2}=-\frac{0.3}{\alpha}\rho\nu R\frac{d\Omega}{dR}
\end{equation}
Replacing these quantities, and applying relations $\dot{M}=3\pi\Sigma\nu$ (valid for a steady accretion), and $n=1/2$ (an acceptable value for advection-dominated discs (see, e.g. \citet{2013MNRAS...434..2275})), one may obtain
\begin{equation}
R>(3h^{2}\alpha\frac{M}{\dot{M}}\sqrt{1.03h^{2}GM})^{2/3}=R_{ins}.
\label{17}
\end{equation}

Finally, we are in need of local magnetic Jeans length, $\lambda_{jmag}$, which takes the form
\begin{equation}
\lambda_{jmag}=\lambda_{j}\sqrt{1+\frac{v_{A}^{2}}{c_{s}^{2}}}=c_{s}\sqrt{\frac{\pi}{G\rho}}\sqrt{1+\frac{v_{A}^{2}}{c_{s}^{2}}}.
\label{18}
\end{equation}
Regarding all above considerations, we will have
\begin{equation}
\lambda_{jmag}=\sqrt{\frac{3.1h^{5}\pi^{2}\alpha M\sqrt{GMR_{ins}}}{\dot{M}}}.
\label{19}
\end{equation}

Hereby, we have our approximately estimated values for $\Sigma_{0}$ and $R_{0}$.

\subsection{What about $R_{in}$ and $R_{out}$?}
\label{sec2.2}

The inner radius ($R_{in}$) differs in two cases of boundary conditions we adopted here. In case of zero torque boundary condition, we choose it to be equal to the Innermost Stable Circular Orbit (ISCO), while in zero mass flux boundary condition, in which the magnetic barrier can be taken into account, we considered the magnetospheric radius as the inner radius. This quantity can be defined as (\citet{2006MNRAS...370L..61}) 
\begin{equation}
R_{m}\approx 60\epsilon_{-3}^{2/3}\dot{M}_{1}^{-2/3}M_{3}^{-4/3}\phi_{30}^{4/3}
\label{20}
\end{equation}
where $\phi_{30} \equiv \phi/(10^{30}~cm^{2}G)$, with $\phi$ is the magnetic flux accumulated in the disc inner side and has been taken to be of about $10^{-2}$, that can be supposed a typical value for GRB's central engine activity ( \citet{2017ApJ...845..64}; \citet{2009MNRAS...398..583}). Also, $\epsilon_{-3}=10^{3}\epsilon$, in which $\epsilon$ is a parameter defined to be the ratio of radial velocity of the accreted matter over their free fall velocity inside the magnetospheric radius (\citet{2017ApJ...845..64}). We adopted a value of $10^{-3}$ for $\epsilon$. Indeed, this parameter is of less certainty (for more discussion about this parameter see, e.g., \citet{2016MNRAS...461..1045}, and \citet{2003PASJ...55..L69}). $\dot{M}_{1}$ and $M_{3}$ denote $\dot{M}/1M_{\odot}~s^{-1}$ and $M_{BH}/3M_{\odot}$, respectively.

On the other hand, as time goes, after the prompt phase, the accretion rate declines. \citet{2008MNRAS...390..781} reported that a neutrino cooled accretion disc, in its late time viscous evolution, experiences a decreasing mass accretion rate with a self similar behavior $\dot{M}\propto t^{-4/3}$. Hence, we approximate the late time accretion rate to be
\begin{equation}
\dot{M}=\dot{M_{0}}(\frac{t}{t_{0}})^{-4/3}
\label{21}
\end{equation}
where $\dot{M_{0}}$ is the accretion rate related to a given time, $t_{0}$, specified (arbitrarily) during the late activity phase. We have fixed $\dot{M_{0}}$ to be of the value of $0.04M_{\odot}/s$ at $t_{0}=50s$ after the prompt phase (that is supposed to have an accretion rate of about $0.1-10M_{\odot}/s$, typically). 

In the context of the outer boundary, we consider $R_{out}$ as a radius at which the viscous evolution of clump sets in. To estimate such a radius, we take shear, self-gravity and tidal forces per unit mass, exerted on the clump, as introduced by \citet{2017MNRAS...464..4399}. We then add Lorentz force in order to account for the large scale effects of magnetic field. 

A clump of linear size $l_{cl}$ and mass $M_{cl}$ will be affected by the shear force per unit mass

\begin{equation}
F_{\nu}=\frac{l_{cl}\dot{M}\Omega}{2\pi M_{cl}}
\label{22}
\end{equation}      
in which the relation $\dot{M}=3\pi\nu\Sigma$, valid for a steady state disc, has been considered. 

Given the self-gravity force per unit mass of the clump, $F_{SG}=\frac{GM_{cl}}{l_{cl}^{2}}$, the tidal force due to the central object, $F_{T}=\frac{GM_{BH}}{r^{2}}(\frac{l_{cl}}{r})$ (\citet{2017MNRAS...464..4399}), and Lorenz force per unit mass, $F_{B}=\frac{B_{r}B_{\phi}}{4\pi\Sigma}$  (\citet{2000Phys. Rev...325..83}), viscous spreading of the clump will start roughly when $F_{\nu}+F_{T}+F_{B}>F_{SG}$. It is worth noting that, making use of equation (\ref{7}) and $\dot{M}=3\pi\Sigma\nu$, Lorentz force can be rewritten in the form of $-\frac{9GM\alpha h}{4r^{2}}$. After all, the condition for being viscously evolved reads

\begin{equation}
\frac{GMl_{cl}}{r^{3}}+\frac{l_{cl}\dot{M}}{2\pi M_{cl}}\sqrt{\frac{GM}{r^{3}}}-\frac{9GM\alpha h}{4r^{2}}-\frac{GM_{cl}}{l_{cl}^{2}}>0
\label{23}
\end{equation}   
The solution to this equation can provide us with an upper limit for the outer boundary $R_{out}$. 

\subsection{Issues on bolometric and X-ray luminosities}
\label{sec2.3}

There are some points highly matter to be noticed. It is the relativistic jet  (which can be created through Blandford-Znajek (BZ) mechanism, as the most viable scenario that can efficiently extract the rotational energy of a spinning black hole by a large-scale magnetic field threading the disc (\citet{1977MNRAS...179..433})), that might produce the GRB's X-ray flares. To be more precise, such emissions are subject to the radiative mechanisms through which the jet power is converted into radiation. For the sake of simplicity, we assume the radiative efficiency in the jet to be constant. Therefore, the bolometric luminosity of the flare will be
\begin{equation}
 L_{bol}=f_{rad}L_{acc}
 \end{equation} 
where $f_{rad}$ is the conversion efficiency of energy from mass accretion into radiation that is a constant based on our assumption. We also adopted $f_{rad}\approx0.5$, regarding discussions made by \citet{2012MNRAS...424..524}. However, in case of zero flux boundary condition, i.e. taking the magnetic barrier into account, and consequently, using the radiation efficiency in MAD (Magnetically Arrested Disk (\citet{2003PASJ...55..L69})) model, we adopted a value of $\sim1.0$ for $f_{rad}$, based on discussions made by \citet{2011MNRAS...418..L79} and \citet{2012MNRAS...423..3083}.  

On the other hand, we need to regard the X-ray bandpass ($0.3-10 keV$), in order to evaluate this model through making a comparison with observations that are performed in a finite X-ray energy band. Thus, an efficiency coefficient is required to consider the X-ray energy band instead of bolometric light-curve. Here, we follow \citet{2010ApJ...724..861} and \citet{2006MNRAS...369..197} to approximate X-ray flare luminosity as a constant coefficient of bolometric luminosity, $L_{X}=f_{X}L_{bol}$, where $f_{X}$ (X-ray flare efficiency) is estimated as $\approx0.1$. 

\section{Results}
\label{sec3}
\subsection{Model parameters and spectral properties correlation}
\label{sec3.1}

\begin{deluxetable*}{cccccccccc}[!b]
	\tablenum{1}
	\tablecaption{Spectral and model parameters for zero torque and zero mass flux boundary condition.\label{tab1}}
	\tablewidth{0pt}
	\tablehead{
		\colhead{$M_{BH}$} & \colhead{$t_{off}$} & \colhead{$h$} & \colhead{$\alpha$} & \colhead{$M_{cl}$} & \colhead{$t_{p}$} & \colhead{$L_{p}$} & \colhead{$\Delta t$} & \colhead{$w$} & \colhead{$k$}\\ \colhead{($M_{\odot}$)} & \colhead{($s$)} & \colhead{{}} & \colhead{{}} & \colhead{($M_{\odot}$)} & \colhead{($s$)} & \colhead{($10^{48}erg/s$)} & \colhead{($s$)} & \colhead{{}} & \colhead{{}} }
	\decimals
	\startdata
	10 & 100.0 & 0.6 & 0.01 & 0.36 & 110.0 & 1.4 & 16.0 & 0.15 & 2.5 \\
	10 & 200.0 & 0.6 & 0.01 & 0.36 & 220.5 & 0.6 & 32.0 & 0.15 & 2.3\\
	10 & 100.0 & 0.5 & 0.01 & 0.25 & 105.0 & 1.6 & 7.5 & 0.07 & 2.7\\
	10 & 200.0 & 0.5 & 0.01 & 0.25 & 210.5 & 0.7 & 15.0 & 0.07 & 2.0\\
	5 & 50.0 & 0.9 & 0.01 & 0.4 & 61.5 & 2.12 & 21.5 & 0.3 & 2.5\\
	5 & 100.0 & 0.9 & 0.01 & 0.4 & 122.5 & 0.73 & 43.0 & 0.3 & 2.7\\
	3 & 50.0 & 0.9 & 0.01 & 0.2 & 57.0 & 2.1 & 12.5 & 0.2 & 2.51\\
	3 & 100.0 & 0.9 & 0.01 & 0.2 & 113.5 & 0.7 & 26.0 & 0.22 & 2.7\\
	3 & 50.0 & 0.9 & 0.1 & 0.76 & 57.0 & 2.7 & 12.5 & 0.2 & 2.5\\
	3 & 100.0 & 0.9 & 0.1 & 0.76 & 114.5 & 1.15 & 26.5 & 0.23 & 2.7\\
	2.5 & 1000.0 & 0.6 & 0.01 & 0.09 & 1094.5 & 0.023 & 148.0 & 0.14 & 2.4\\
	2.5 & 1000.0 & 0.5 & 0.01 & 0.06 & 1048.5 & 0.03 & 69.0 & 0.07 & 2.2\\
	2.5$^{*}$ & 200.0 & 0.8 & 0.15 & 0.6 & 215.0 & 6.0 & 26.0 & 0.12 & 2.7\\ 
	2.5$^{*}$ & 300.0 & 0.8 & 0.15 & 0.6 & 340.5 & 1.12 & 70.5 & 0.2 & 2.7\\ 
	3$^{*}$ & 200.0 & 0.8 & 0.13 & 0.7 & 218.0 & 8.0 & 31.5 & 0.15 & 2.7\\
	3$^{*}$ & 300.0 & 0.8 & 0.13 & 0.7 & 349.0 & 1.5 & 85.0 & 0.24 & 2.6\\
	\enddata
	\tablenotetext{*}{Cases with zero flux boundary condition.}
\end{deluxetable*}

Our model consists of some parameters, $M_{BH}$, $h$, and $\alpha$, whose effects on shape parameters are important to be noticed. We introduce two shape parameters $k$ and $w$ to be the '$asymmetry~parameter$', and the ratio of the width, $\Delta t=t_{2}-t_{1}$, to the peak time, $t_{p}$, respectively. $t_{p}$ is the time at which the peak luminosity $L_{p}$ is expected. $t_{1}$ and $t_{2}$ refer to the times with fluxes of around $\frac{1}{e}L_{p}$ (the former during the rise time interval and the latter during the decay one). The asymmetry parameter is regarded to be the ratio $t_{d}/t_{r}$, with rise time $t_{r}=t_{p}-t_{1}$ and decay time $t_{d}=t_{2}-t_{p}$.

To have an intuition of how model parameters affect those related to light curve shape, we provided some data, predicted by our model, in table (\ref{tab1}). It should be mentioned that we considered $0.5\leqslant h\leqslant 1$ (valid for the late time advection dominated phase of GRB's central engine (e.g., \citet{2008MNRAS...390..781})), and $0.01\leqslant \alpha\leqslant 0.2$ which is a physically justified interval (e.g., \citet{2012MNRAS...423..3083}). Generally speaking, we interestingly found the shape parameters, $w$ and $k$, are not significantly sensitive to the model parameters, regardless of some scatterings. The fact that has been confirmed by the statistical analysis of observational data, and will be discussed in more details in section (\ref{sec3.2}).

To elaborate how effective the model parameters are, we point to their impact on the clump mass and, subsequently, on the maximum amount of flux and width parameter, which might directly affect the light-curve shape. First, as $h$ grows, the clump mass increases, and this leads to a decline in the peak luminosity, and an increase in width parameter as well as peak time, i.e. the light-curve gets wider with less maximum radiated flux. At first glance, it might appear to contradict the fact that the larger the clump mass is, the higher the total radiated energy gets ($E\propto M_{cl}$). For one thing, we found in the lower case of $h$, the radius at which the viscous evolution is triggered, is less than the thicker clump. For instance, $R_{0}$ (that is chosen to be the same as $R_{ins}$, since the viscous evolution constraint (\ref{23}) is respected by this radius, in this case) is $905.2R_{g}$ for the set of parameters $M_{BH}=10,h=0.5,\alpha=0.01$ and $t_{off}=200s$, while it is of about $1303.5R_{g}$ in the similar case with $h=0.6$. For another, in the former the clump is denser than the latter. Subsequently, the flare duration (which might be estimated to be of the order of the width parameter) for the thinner clump will be considerably less than the thicker one, as one can see in data provided in table (\ref{tab1}). Thus, we are of the opinion that the heavier clump is expected to radiate more total energy, although, its lower density located at a farther distance causes such a decline in the luminosity. On the other hand, this behavior can be explained through the explicit dependency of luminosity on $h$. Concerning equation (\ref{4}), clump mass increases by $h^{2}$. Also, $R_{0}$ ($\sim R_{ins}$) is proportional to $h^{2}$, as appears in equation (\ref{17}). These two proportionalities together with the relation between $l_{cl}$ and $h$ ($l_{cl}(\sim \lambda_{jmag})\propto h^{3}$) cause $\Sigma_{0}$ to be proportional to $h^{-3}$, regarding equation (\ref{14}). Thus, one may conclude that $\Sigma$ has a reverse relation with $h$ (i.e. $\Sigma\propto h^{-1.5(1+1/h)}$), and this subsequently leads luminosity to drop as $h$ grows, considering equation (\ref{11}). Apparently, this result opposes \citet{2017MNRAS...464..4399} outcome, as they conclude that for the same initial radius ($R_{0}$), a more massive clump produces a larger peak luminosity. Such an inconsistency emanates from ignoring correlations we considered between model parameters in our framework, and fixing them arbitrarily.

Secondly, our data demonstrate that any increase in $\alpha$ parameter might enhance remarkably the clump mass, while the width parameter might not be affected considerably, so that the peak luminosity grows. The fact that has been also inferred by \citet{2017MNRAS...464..4399}. Finally, we came into conclusion that a black hole mass growth may lead to a heavier clump with a rather higher luminosity that happens in a longer duration time scale. In this case, the clump mass growth is apparently more remarkable than duration time increase, so that a final higher luminosity is provided. 

Over all, such correlations between model parameters and spectral quantities are inferable from our data, however, in agreement with \citet{2017MNRAS...464..4399} and observational analysis, the shape parameters $w$ and $k$ are not strongly affected by our model parameters, regardless of their somehow scattered behavior.

\subsection {Observations and model predictions}
\label{sec3.2} 

From the study of 113 flares in the X-ray Telescope (XRT) $0.3-10 keV$ energy band, and four subenergy bands, proceeded by \citet{2010MNRAS...406..2113}, some observational characteristics of the X-ray light curves can be clarified as follows

\begin{figure}[!b]
	\centering
	\includegraphics[width=90mm]{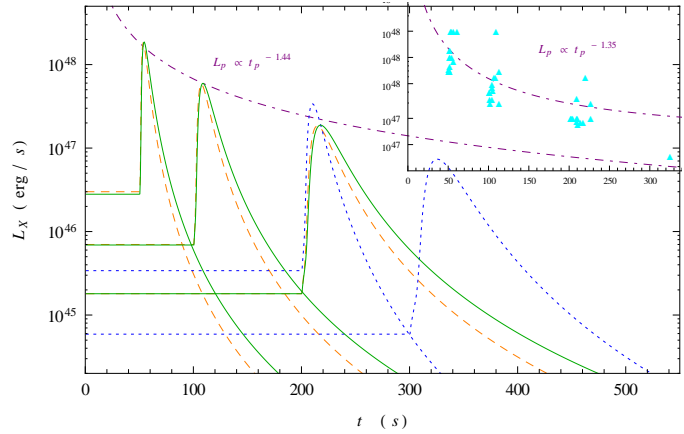}
	\caption{\small{X-ray luminosity light curve has been plotted for three sets of model parameters: $M_{BH}=3M_{\odot},~h=0.8$ and $\alpha=0.13$ with curves of green color, $M_{BH}=2.5M_{\odot},~h=0.8$ and $\alpha=0.15$ with orange dashed curves, and $M_{BH}=5M_{\odot},~h=0.6$ and $\alpha=0.22$ with dotted curves of blue color. The vertical axis is in logarithmic scale. Four different offset times, $t_{off}=50s,100s$, $200s$ and $300s$, in case of zero torque boundary condition, are included. The dot-dashed lines in purple, depicts the best power law fit for the peak luminosity and peak time, which are in a good agreement with observations. } }
	\label{fig1}
\end{figure} 

\begin{figure}
	\includegraphics[width=90mm]{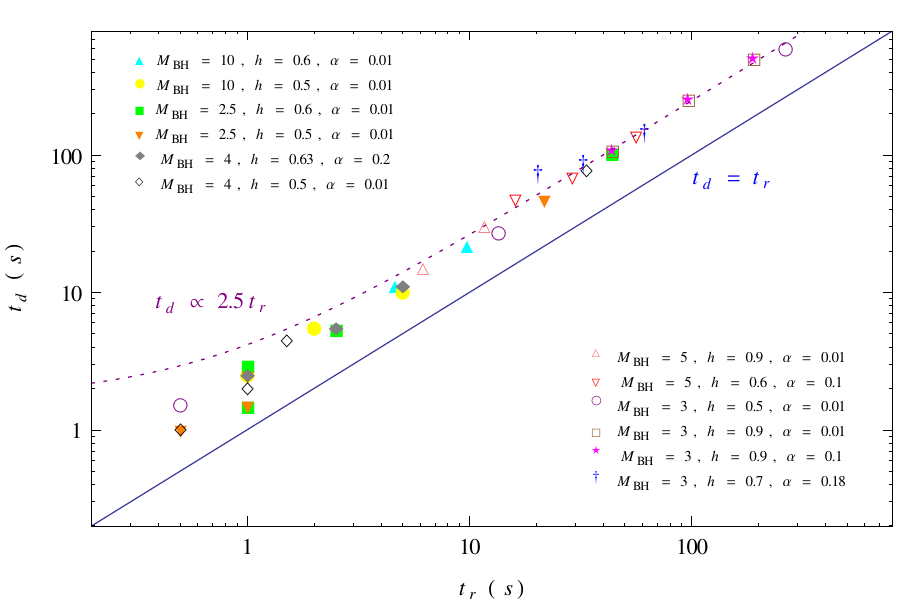}
	\includegraphics[width=90mm]{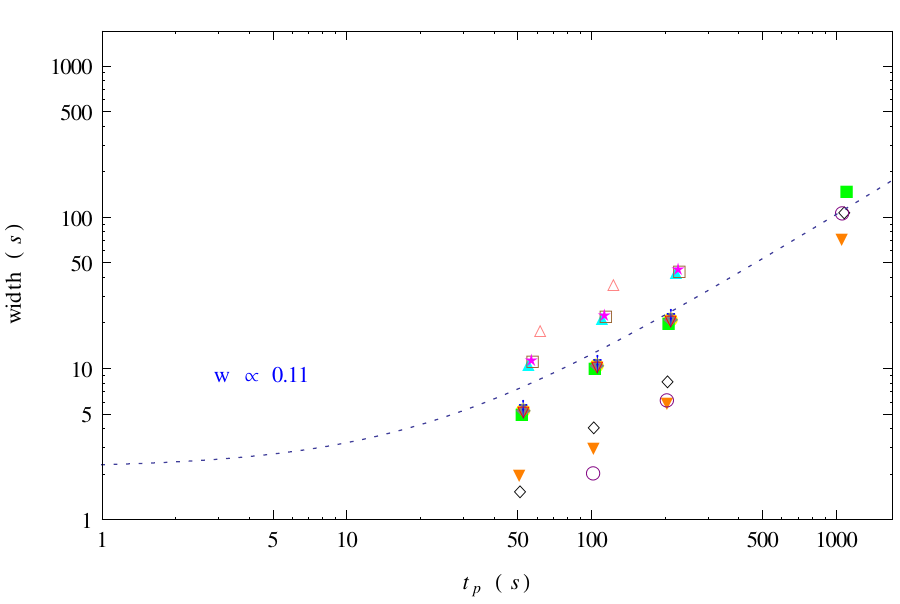}
	\caption{\small{Decay time versus rise time, for various sets of model parameters and offset times, is depicted in the left panel with its linear fit in dotted purple line. The right one is the best linear fit for the width parameter. Both fits show a rather good agreement with observations as discussed in the text. Zero torque boundary condition has been considered.}}
	\label{fig2}
\end{figure}

(i) The rise over decay time ratio is constant, implying that both timescales grow by the same factor, so that $t_{d}\approx 2t_{r}$. Consequently, flares are self-similar in time.

(ii) The width linearly evolves with the peak time: $w\approx 0.2$. These two points are the key features that strongly distinguish the flare emission from the prompt phase.

Moreover, analyzing 468 bright X-ray flares from the GRBs observed by Swift between 2005 and 2015, \citet{2016ApJS...224..20} argued that the peak luminosity decreases with the peak time, following a power-law behavior $L_{p}\propto t_{p}^{-1.27}$.

In general, for both boundary conditions, our model leads to a rather observationally well matched predictions, considering the mentioned shape parameters.

\begin{figure}[!b]
	\centering
	\includegraphics[width=90mm]{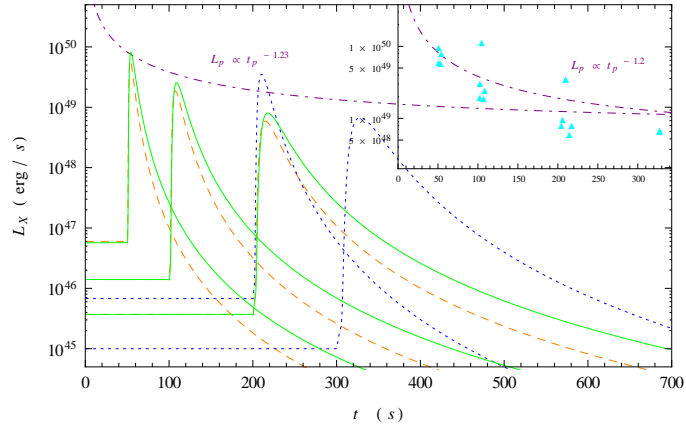}
	\caption{\small{The same as figure (\ref{fig1}) in case of zero flux boundary condition. Considering all sets of model parameters (those of figure (\ref{fig4}) have been also taken into account), the fit for the inner plot complies with observations as well as the other one (with only three sets of model parameters).  }}
	\label{fig3}
\end{figure} 

\begin{figure}
	\includegraphics[width=90mm]{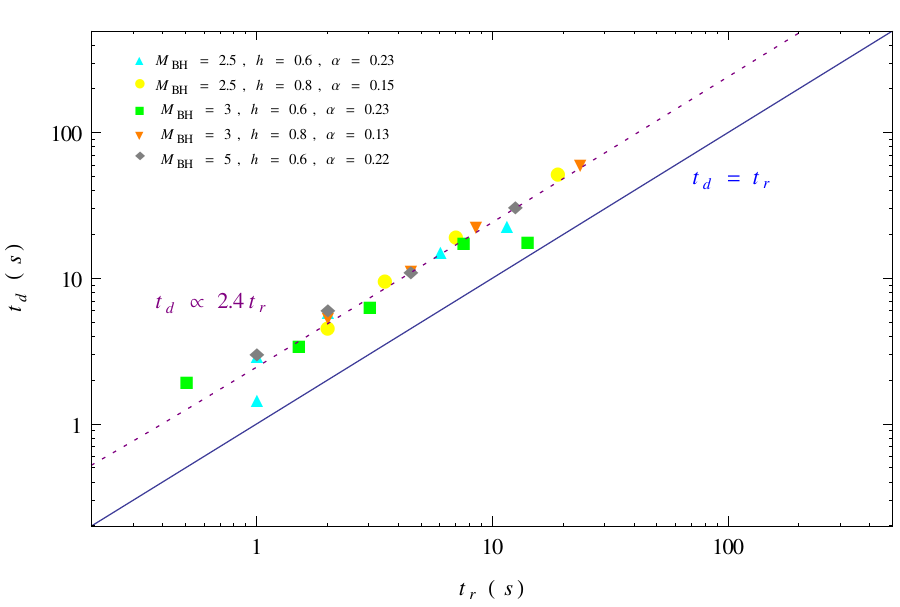}
	\includegraphics[width=90mm]{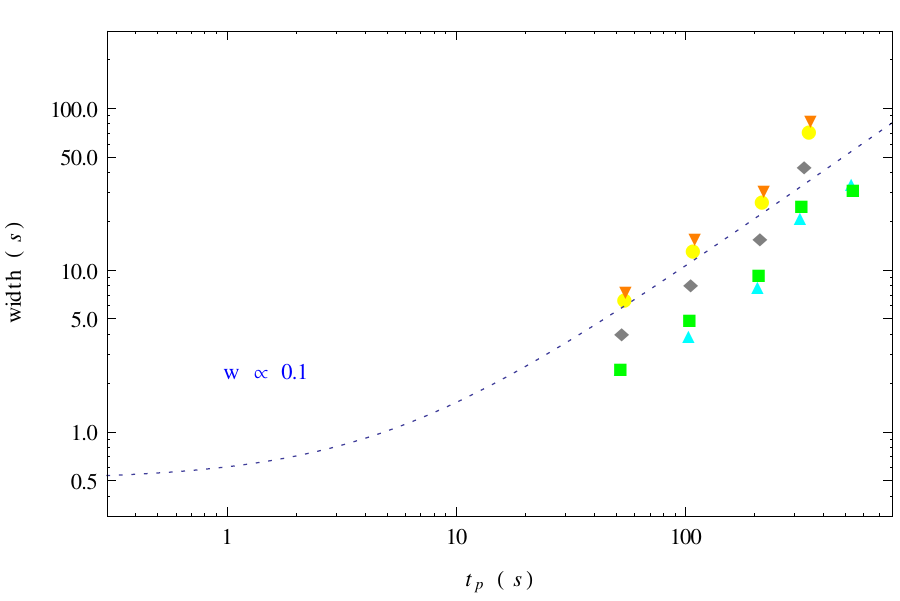}
	\caption{\small{Fits for asymmetry parameter (the left panel), and width to peak time ratio (the right one) is displayed with zero flux boundary condition has been regarded. }}
	\label{fig4}
\end{figure}

In case of zero torque boundary condition, figures (\ref{fig1}) and (\ref{fig2}) demonstrate how our model respects these three criteria. We have considered a variety of model parameters, as declared inside panels, and different offset times during the early flaring activity phase (with $t_{p}<1000s$), i.e. $50s, 100s, 200s$ and $1000s$, in order to enhance the validity of our fits.

Figure (\ref{fig1}) reflects the logarithmic behavior of X-ray luminosity in time. Four different offset times $50s, 100s $, $200s$ and $300s$ have been regarded. The best fit for peak luminosity versus peak time, the dot-dashed line in purple, indicates a power law trend with a power index of $\sim -1.44$, which is interestingly close to the value discussed by \citet{2016ApJS...224..20}, i.e.  $-1.27$. These light curves have been plotted for three sets of model parameters: $M_{BH}=3M_{\odot},~h=0.8$ and $\alpha=0.13$ with curves of green color, $M_{BH}=2.5M_{\odot},~h=0.8$ and $\alpha=0.15$ with orange dashed curves, and $M_{BH}=5M_{\odot},~h=0.6$ and $\alpha=0.22$ with dotted curves of blue color. The panel inside shows the fit for a broader range of model parameters including those considered in figure (\ref{fig2}). Being of a rather similar power index, i.e. $\sim -1.35$, the variation of model parameters does not affect the correlation we obtained in our model which is in a good agreement with its phenomenological relation, namely $L_{p}\propto t_{p}^{-1.27}$. On the other hand, as expressed in section (\ref{sec3.1}), figure (\ref{fig2}) reveals the robustness of the light curve shape in our model through the considerable (acceptable) insensitivity of the asymmetry (width over peak time) parameter. Furthermore, our predicted values for these two shape parameters, i.e. $k\approx 2.5$ and $w\approx 0.11$, are in a rather good agreement with what predicted by \citet{2010MNRAS...406..2113}, in which $k\approx 2$ and $w\approx 0.2$.   

In order to get the results for zero mass flux boundary condition and, consequently, take the magnetic barrier into account, we need to probe for set of model parameters which satisfy the condition

\begin{equation}
\tau=\frac{t_{\nu}}{t_{diff}}<1 
\label{25}
\end{equation}  
where $t_{diff}\approx \frac{H}{v_{A}}$ is diffusion time scale that estimates the magnetic field buoyancy and its rising time toward the disc surface (\citet{2017ApJ...845..64}). Such a limitation assures us magnetic barrier gets probable during the clump's accretion due to the accumulation of the magnetic flux inside the inner regions (\citet{2017ApJ...845..64}; \citet{2009MNRAS...398..583}). The adopted parameters in figures (\ref{fig3}) and (\ref{fig4}) comply this criteria. For instance, regarding $M_{BH}=2.5M_{\odot}, h=0.8$ and $\alpha=0.15$ we found an average value of about $0.8$ for $\tau$.

Figure (\ref{fig3}) is an illustration of X-ray luminosity variation versus time for different sets of model parameters, and offset times $50s, 100s, 200s $ and $300s$. These parameters are similar to those considered in figure (\ref{fig1}), and, to show the generality of our result, we extended sets of parameters for the inner panel to include those regarded in figure (\ref{fig4}). Besides the fact that the power law fit for $L_{p}$ and $t_{p}$, has a good agreement with phenomenological predictions, taking the magnetic barrier into account leads to an order of magnitude increase in the peak luminosity in comparison with the similar cases in zero torque boundary condition. This is an expected result according to general relativistic magnetohydrodynamic simulations performed, e.g., by \citet{2012MNRAS...423..3083} and \citet{2011MNRAS...418..L79}.

Regarding figure (\ref{fig4}), our fits for the other two shape parameters $k$ and $w$, confirm a satisfactory robustness of our model, in case of zero mass flux boundary condition, to reproduce the light curve shape, as well as in zero torque boundary condition.      

However, on the subject of magnetic barrier mechanism we must point out some worth mentioning issues. On the one hand, \citet{2014ApJ...791..69}, through the study of a self-gravitating nuetrino dominated accretion disc (called NDAF), confirmed that instabilities appear in the outer parts and away from the equatorial plane. From a magnetic point of view, \citet{2017ApJ...845..64} probed for the possibility of fragmentation in a self-gravitating magnetized NDAF and found it possible in the outer and off-equatorial plan regimes, as well. Thus, we think that clumps might be thinner than what we expected from the late time advection dominated central engine's vertical structure, namely smaller values of $h$ parameter should be of more validity to be considered. As a consequence, clumps of smaller masses would be of more probability to form. On the other hand, regarding $\alpha$ in its physically meaningful range, we found that magnetic barrier is mainly plausible in case of heaver clumps (with masses in the order of $\sim(0.4-0.8)M_{\odot}$). Therefore, we think such a limitation might cause magnetic barrier to get less probable to govern the evolution of the generated clumps. Moreover, we considered $\phi$ to be of the order of $10^{-2}$, which might be an over estimation for the late time accumulated magnetic flux. Since as time progresses and accretion rate drops, both clump's magnetic field and accumulated magnetic flux decline. Such a drop in magnetic field, which gets more probable for higher values of $t_{off}$, might weaken the possibility of magnetic barrier even more (see, e.g., \citet{2012MNRAS...423..3083} and references therein).

In the context of power-law correlation between $L_{p}$ and $t_{p}$, there are some points of importance to be noticed. We need to emphesize that the time dependency we considered to set our model parameters such as the radius at which clump is created, can not lead to such a power-law behavior. Since we have used these relations to make some physically justified constraints on our model parameters, like $\dot{M}$ and $R_{0}$, and these relations have not been applied directly in the time evolution equation of luminosity. In particular, equation (\ref{21}) has been only used to determine the accretion rate at $t_{off}$ and not been considered in the main formalism of the model, that provides us with the time evolution of luminosity. In other words, through the use of these relations we only aimed to fix some model parameters regarding their offset time at which the clump starts to accret. Hence, their time dependencies have not influenced the clump's evolution directly, so that one can think of this power-law trend as a general outcome of our model.

\section{Summery, conclusions and discussion}
\label{sec4}

We reconstructed the quantitative framework established by \citet{2017MNRAS...464..4399}, through a magnetic style in which small and large scale effects of magnetic field have been accounted. To this end, we added one more term corresponding to the torque exerted by Lorentz force (see equation (\ref{5})). To solve this evolution equation, some approximations and assumptions have been implemented in order to be able to adopt a similar semi-analytic approach as proposed by \citet{2011MNRAS...410..1007} and followed by \citet{2017MNRAS...464..4399}. Through this method, we have been provided by final solutions which differ from those achieved by the previous work, in some aspects. First, for both cases of boundary conditions, the power law dependency on radius changed in comparison with solutions obtained by \citet{2011MNRAS...410..1007}. Second, considering $h$ as an arbitrary parameter not to be fixed only at "$1$" (this affects $t_{\nu_{in}}$ in the final solution) causes another difference in our result with respect to what has been achieved by \citet{2017MNRAS...464..4399}. Finally, to estimate model parameters such as clump mass, we limited ourselves to the correlations introduced between these parameters, so that our results are able to verify the framework we defined, however, in \citet{2017MNRAS...464..4399} such correlations have been of less importance to be considered in providing data. On the other hand, to investigate the possibility of magnetic barrier occurrence during each clump's viscous evolution, we made a comparison between diffusion and viscous time scales. Moreover, two different boundary conditions have been imposed, i.e. zero torque and zero mass flux boundary conditions, in the hope to distinguish the presence of the accumulated magnetic flux inside the inner regions, and consequently, magnetic barrier mechanism. We then studied the validity of our model through some key analogies to phenomenological findings. Our main conclusions can be identified as follows:

\begin{enumerate}
	\item The ratio of width over peak time found to be an almost constant, although, some how scattered, parameter of an average value $\sim0.1$, which is in a rather good agreement with its phenomenological estimate, $\sim0.2$. We also came into conclusion that magnetic barrier, if happens, will not change this parameter significantly, instead, it might cause our data to get less scattered (left panel in figure (\ref{fig4})). 
	\item The model parameter insensitivity of X-ray light curves' skewness (the asymmetry parameter), reveals another robust aspect of our model. This important feature is respected in case of both boundary conditions, with a similar value of about $\sim2.4$, which has a satisfying agreement with its observational prediction, i.e. $\sim2$. 
	\item Interestingly, the power law correlation between the maximum value of luminosity and peak time, with a value of about $\sim-1.3$, found to be close to what has been estimated by \citet{2016ApJS...224..20} from phenomenological point of view, i.e. $\sim-1.27$. Both boundary conditions adopted here, respect this feature.
\end{enumerate}

There are some more points worth discussing, here. For one thing, \citet{2017MNRAS...464..4399} found the bolometric luminosity to be far from being satisfactory in comparison to observations. However, after conducting a spectral correction through the adoption of a time dependent X-ray efficiency, they modified the results to be well-matched with observations. Whereas, considering the large scale effect of magnetic field led us to obtain shape parameters in a close agreement with phenomenological values, in both cases of bolometric and X-ray luminosities. Note that, in our study, bolometric luminosity only differs by a constant coefficient from X-ray luminosity, $L_{X}=f_{X}L_{bol}$. Thus, the bolometric and X-ray light curves are of the same shape parameters in our model.

For another, we assumed $f_{rad}$ and $f_{X}$ to be constants. Accounting for any inconstancy about these two parameters might affect the light curve shape. On the one hand, simulations make us to conclude that the efficiency in converting torus mass into jet energy varies from a few percent up to more than 100\%. The black hole spin, the disc thickness, and the magnetic flux might effectively alter this parameter (see, e.g., \citet{2013ApJ...762L..18} and references therein). On the other hand, regarding $f_{X}$, \citet{2017MNRAS...464..4399} argued that a time dependent X-ray efficiency with a decreasing trend in time (which is in agreement with the 'curvature effect', if the emitting region has an accelerated bulk relativistic motion (\citet{2015ApJ...808..33},\citet{2016ApJ Letters...824..L16})) might affect the spectral properties of X-ray flares. 

Separate from these considerations, together with the other probable modifications, our simple model could efficiently reproduce X-ray flare's spectral properties. However, more precise studies via simulations appear to be a crucial need to lighten the dark nature of late time evolution of GRB's central engine.

\appendix

\section{Boundary Conditions and Green's Functions}
\label{appA}

 Following \citet{2017MNRAS...464..4399}, to find the Green function, we adopted the viscosity to respect a radial power law, $\nu\propto R^{n}$, and assumed a separable ansatz of the form $\Sigma (R,t)=R^{p}\sigma (R) e^{-\lambda t}$, where $p$ and $\Lambda$ are real numbers and $\sigma$ is an arbitrary function of $R$, so that equation (\ref{9}) can be rewritten as a Bessel differential equation (\citet{2011MNRAS...410..1007})  
 \begin{eqnarray}
 \frac{\partial^{2}\sigma}{\partial R^{2}}+\frac{\partial\sigma}{R\partial R}(1.5+3/h+2(n+p))\nonumber\\\nonumber\\
 +\sigma (R)[\frac{\Lambda}{3s}R^{-n}+\frac{(n+p)(0.5+3/h+(n+p))}{R^{2}}]=0
 \label{A1}
 \end{eqnarray}
 where $s=\nu R^{-n}$ is a constant. It should be mentioned that we considered the Newtonian potential with $\Omega=\sqrt{\frac{GM}{R^{3}}}$. We also choose $p=n+\frac{1-b_{1}}{2}$, with $b_{1}=1.5+3/h$, and $\Lambda=3sk^{2}$. All these considerations besides comparing equation (\ref{A1}) with the transformed Bessel function 
 \begin{equation}
 \frac{\partial^{2}y}{\partial x^{2}}-\frac{\partial y}{\partial x}(\frac{2\alpha-1}{x})+y(x)[\frac{\alpha^{2}-l^{2}\gamma^{2}}{x^{2}}+\beta^{2}\gamma^{2}x^{2\gamma-2}]=0
 \label{A2}
 \end{equation}
 yield
 \begin{equation}
 \gamma=1-n/2; ~\beta=\frac{k}{1-n/2}; ~\alpha=2n~\&~l=\frac{\mid1-b_{1}\mid}{2(1-n/2)}\nonumber
 \end{equation}
 Now, the solution to the equation (\ref{10}) can be written in the following form
 \begin{equation}
 \sigma_{k}(R)=R^{-2n}[A(k)J_{l}(ky)+B(k)Y_{l}(ky)]
 \label{A4}
 \end{equation}
 where $y=\frac{R^{1-n/2}}{1-n/2}$. It is worth noting that in case of non-integer $l$, $Y_{l}$ must be replaced by $J_{-l}$. Integrating the above solution over all possible $k$-modes gives the solution
 \begin{equation}
 \Sigma (R,t)=\int_{0}^{\infty}R^{-n-1/4}[A(k)J_{l}(ky)+B(k)Y_{l}(ky)]e^{-3sk^{2}t} dk
 \label{A5}
 \end{equation}
 The mode-weighting functions $A(k)$ and $B(k)$ are determined by the boundary conditions and the initial surface density profile $\Sigma (R,t=0)$.

We just elaborate the zero mass flux boundary condition, here, as the zero torque one is implemented in a similar way. Before any further calculation to get the final solution, it is better to determine $s$ in the power low expression for viscosity as follows.
The (magnetic) viscosity can be parametrized as $\nu=\alpha c_{s}^{2}/\Omega$ (\citet{1973A&A...24..337}; \citet{1991MNRAS...248..754}). On the other hand, invoking hydrostatic equilibrium perpendicular to the disc plane gives us $H=\frac{c_{s}}{\Omega}$, and, consequently, one can infer $\nu=\alpha h^{2}\sqrt{GM}R^{1/2}$, which yields $s=\alpha h^{2}\sqrt{GM}$ and $n=1/2$. 
Furthermore, regarding the continuity equation 
\begin{equation}
R\frac{\partial\Sigma}{\partial t}+\frac{\partial\Sigma R v_{R}}{\partial R}=0,
\label{A6}
\end{equation}
together with 
\begin{equation}
\dot{M}=-2\pi\Sigma R v_{R},\nonumber
\end{equation}
lead us to
\begin{equation}
\frac{\partial\Sigma}{\partial t}=\frac{1}{2\pi R}\frac{\partial\dot{M}}{\partial R}.
\label{A8}
\end{equation}
Comparing equations (\ref{A8}) and (\ref{9}) provides us with the following result
\begin{equation}
\dot{M}\propto[\frac{\partial}{\partial R}(R^{3}\nu\Sigma \frac{d\Omega}{dr})+3\frac{d\Omega}{dR}\frac{\Sigma}{h}\nu R^{2}]|_{R_{in}}=0.
\end{equation}

By a substitution of the corresponding expressions for $\nu$, $\Omega$ and $\Sigma$, and making use of the recurrence relations between different Bessel functions
\begin{eqnarray}
J_{l-1}(x)-J_{l+1}(x)=2J^{\prime}_{l}(x)\nonumber\\\nonumber\\
J_{l-1}(x)+J_{l+1}(x)=2l\frac{J_{l}}{x}(x),\nonumber
\end{eqnarray}
the following relation can be achieved
\begin{equation}
\frac{A(\kappa)}{B(\kappa)}=-\frac{Y_{l-1}(\kappa)}{J_{l-1}(\kappa)},
\label{A9}
\end{equation}
where $\kappa=ky_{in}$.
As goes below, we need to obtain some more ingredients to get the final solution for $\Sigma (R,t)$. First, 
\begin{equation}
\Lambda=3sk^{2}=3\nu R^{-n}k^{2},\nonumber
\end{equation}
and, secondly, we know
\begin{equation}
t_{\nu}=\frac{2}{3}\frac{R^{2}}{\nu}=\frac{2}{3}\frac{R^{2-n}}{s}.
\label{A11}
\end{equation}
Thus, $\Lambda$ can be rewritten in the following form
\begin{equation}
\Lambda=-2\kappa^{2}(1-n/2)^{2}\frac{1}{t_{\nu_{in}}}.\nonumber
\end{equation}
Imposing all above achievements, the expression for surface density takes the form
\begin{equation}
\Sigma=\int_{0}^{\infty}R^{-n+(1-b_{1})/2}[c(\kappa)\kappa^{-1}][Y_{l-1}(\kappa)J_{l}(\kappa x)-Y_{l}(\kappa x)J_{l-1}(\kappa)]e^{-2\kappa^{2}(1-n/2)^2t/t_{\nu_{in}}}\kappa d\kappa\nonumber
\end{equation} 
with $t_{\nu_{in}}$ is viscous timescale of the inner radius of the disc, and $x=y/y_{in}$. The mode weight $c(\kappa)$ might be obtainable by making use of generalized Weber transform (\citet{2011MNRAS...410..1007}; \citet{2007ApJ...642..354})
\begin{equation}
\phi_{l}(x)=\int_{0}^{\infty}\frac{W_{l}(k,x;a,b)}{Q_{l}^{2}(k,;a,b)}\Phi_{l}(\kappa)\kappa d\kappa\nonumber\\\nonumber\\
\Phi_{l}(\kappa)=\int_{1}^{\infty}W_{l}(\kappa,x;a,b)\phi_{l}(x)xdx
\end{equation}
where
\begin{equation}
W_{l}=J_{l}(\kappa x)[(a-lb)Y_{l}(\kappa)+b\kappa Y_{l-1}(\kappa)]-Y_{l}(\kappa x)[(a-lb)J_{l}(\kappa)+b\kappa J_{l-1}(\kappa)]\nonumber
\end{equation}
and
\begin{equation}
Q_{l}^{2}=[(a-lb)Y_{l}(\kappa)+b\kappa Y_{l-1}(\kappa)]^{2}-[(a-lb)J_{l}(\kappa)+b\kappa J_{l-1}(\kappa)]^{2}\nonumber
\end{equation}
If $a=1$ and $b=0$, the pair is identical to the ordinary Weber transform. However, $a=l$ and $b=1$ correspond to our desired boundary condition, i.e. zero mass flux. After all, what would be the Green function?
Clearly,
\begin{equation}
R^{n-(1-b_{1})/2}\Sigma (R,t=0)=\int_{0}^{\infty}[c(\kappa)\kappa^{-1}][Y_{l-1}(\kappa)J_{l}(\kappa x)-Y_{l}(\kappa x)J_{l-1}(\kappa)]\kappa d\kappa.\nonumber
\end{equation}  
Finally, through applying the above considerations and running some more calculations, one may achieve the following result for $c(\kappa)$,
\begin{equation}
c(\kappa)=\int_{1}^{\infty}\frac{R^{n-(1-b_{1})/2}\Sigma (R,0)[Y_{l-1}(\kappa)J_{l}(\kappa x)-Y_{l}(\kappa x)J_{l-1}(\kappa)]xdx}{[Y^{2}_{l-1}(\kappa)+J_{l-1}^{2}(\kappa)]}\nonumber
\end{equation} 
Thus,
{\small \begin{eqnarray*}
		\Sigma (R,t)=\int_{0}^{\infty}(\frac{\int_{r_{in}}^{\infty}[Y_{l-1}(\kappa)J_{l}(\kappa x^{\prime})-Y_{l}(\kappa x^{\prime})J_{l-1}(\kappa)]\Sigma (R^{\prime},0)\frac{(R^{\prime})^{(1+b_{1})/2}}{R_{in}^{2-n}}dR^{\prime}}{[Y^{2}_{l-1}(\kappa)+J_{l-1}^{2}(\kappa)]})(1\nonumber\\\nonumber\\
		-n/2)R^{-n+(1-b_{1})/2}[Y_{l-1}(\kappa)J_{l}(\kappa x)-Y_{l}(\kappa x)J_{l-1}(\kappa)]e^{-2\kappa^{2}(1-n/2)^2t/t_{\nu_{in}}}\kappa d\kappa
	\end{eqnarray*} }
	Considering equation (\ref{10}), Green's function would be obtainable as
	{\small \begin{eqnarray}
		G(R,R^{\prime},t)=\int_{0}^{\infty}\frac{[Y_{l-1}(\kappa)J_{l}(\kappa x^{\prime})-Y_{l}(\kappa x^{\prime})J_{l-1}(\kappa)]}{[Y^{2}_{l-1}(\kappa)+J_{l-1}^{2}(\kappa)]}\frac{(R^{\prime})^{(1+b_{1})/2}}{R_{in}^{2-n}}(1\nonumber\\\nonumber\\
		-n/2)R^{-n+(1-b_{1})/2}[Y_{l-1}(\kappa)J_{l}(\kappa x)-Y_{l}(\kappa x)J_{l-1}(\kappa)]e^{-2\kappa^{2}(1-n/2)^2t/t_{\nu_{in}}}\kappa d\kappa
		\end{eqnarray} }

	Finally, regarding the initial surface density as
	\begin{equation}
	\Sigma(R,t=0)=\Sigma_{0}R_{0}\delta(R-R_{0})
	\end{equation}
	in which $\Sigma_{0}$ and $R_{0}$ are arbitrary (\citet{2011MNRAS...410..1007}), we have
	{\small \begin{eqnarray}
		\frac{\Sigma}{\Sigma_{0}}=\int_{0}^{\infty}\frac{R}{R_{0}}^{-n+(1-b_{1})/2}(\frac{R_{0}}{R_{in}})^{2-n}(1-n/2)e^{-2\kappa^{2}(1-n/2)^2t/t_{\nu_{in}}}[Y_{l-1}(\kappa)J_{l}(\kappa x_{0})\nonumber\\\nonumber\\
		-Y_{l}(\kappa x_{0})J_{l-1}(\kappa)]\frac{[Y_{l-1}(\kappa)J_{l}(\kappa x)-Y_{l}(\kappa x)J_{l-1}(\kappa)]}{[Y^{2}_{l-1}(\kappa)+J_{l-1}^{2}(\kappa)]}\kappa d\kappa
		\label{A22}
		\end{eqnarray}}

Likewise, in case of zero central torque boundary condition, one can evaluate Green's function as

	{\small \begin{eqnarray}
		G(R,R^{\prime},t)=\int_{0}^{\infty}\frac{[Y_{l}(\kappa)J_{l}(\kappa x^{\prime})-Y_{l}(\kappa x^{\prime})J_{l}(\kappa)]}{[Y^{2}_{l}(\kappa)+J_{l}^{2}(\kappa)]}\frac{(R^{\prime})^{(1+b_{1})/2}}{R_{in}^{2-n}}(1\nonumber\\\nonumber\\
		-n/2)R^{-n+(1-b_{1})/2}[Y_{l}(\kappa)J_{l}(\kappa x)-Y_{l}(\kappa x)J_{l}(\kappa)]e^{-2\kappa^{2}(1-n/2)^2t/t_{\nu_{in}}}\kappa d\kappa
		\end{eqnarray} } 
and, subsequently, the surface density will read

	{\small \begin{eqnarray}
		\frac{\Sigma}{\Sigma_{0}}=\int_{0}^{\infty}\frac{R}{R_{0}}^{-n+(1-b_{1})/2}(\frac{R_{0}}{R_{in}})^{2-n}(1-n/2)e^{-2\kappa^{2}(1-n/2)^2t/t_{\nu_{in}}}[Y_{l}(\kappa)J_{l}(\kappa x_{0})\nonumber\\\nonumber\\
		-Y_{l}(\kappa x_{0})J_{l}(\kappa)]\frac{[Y_{l}(\kappa)J_{l}(\kappa x)-Y_{l}(\kappa x)J_{l}(\kappa)]}{[Y^{2}_{l}(\kappa)+J_{l}^{2}(\kappa)]}\kappa d\kappa
		\label{24}
		\end{eqnarray}}

\end{document}